# Force-distance studies with piezoelectric tuning forks below 4.2K

J. Rychen, T. Ihn, P. Studerus, A. Herrmann, K. Ensslin

*Solid State Physics Laboratory, ETH Zurich, CH-8093 Zurich, Switzerland*

H. J. Hug, P.J.A. van Schendel, H.J. Güntherodt

*Institute of Physics, University of Basel, CH-4056 Basel, Switzerland*

(Last modified 9 September 1999 9:33 am)

**Abstract**

Piezoelectric quartz tuning forks have been employed as the force sensor in a dynamic mode scanning force microscope operating at temperatures down to 1.7 K at He-gas pressures of typically 5 mbar. An electrochemically etched tungsten tip glued to one of the tuning fork prongs acts as the local force sensor. Its oscillation amplitude can be tuned between a few angstroms and tens of nanometers. Quality factors of up to 120000 allow a very accurate measurement of small frequency shifts. Three calibration procedures are compared which allow the determination of the proportionality constant between frequency shift and local force gradient based on the harmonic oscillator model and on electrostatic forces. The calibrated sensor is then used for a study of the interaction between the tip and a HOPG substrate. Force gradient and dissipated power can be recorded simultaneously. It is found that during approaching the tip to the sample considerable power starts to be dissipated although the force gradient is still negative, i.e. the tip is still in the attractive regime. This observation concurs with experiments with true atomic resolution which seem to require the same tip-sample separation.

*PACS:*

## 1. Introduction

The interaction of the tip with the surface of a sample is of crucial importance for the operation of scanning force microscopes [1]. A renewed interest in this subject has arisen since atomic resolution was achieved in the dynamic mode on the surfaces of insulators measured in UHV which were previously not accessible for local investigations on the atomic scale [2]. Recently, atomic resolution was achieved in the dynamic mode on highly oriented pyrolytic graphite (HOPG), a so-called van der Waals surface, at temperatures of 10 K [3]. The low temperatures were necessary in order to reduce the thermal noise and provide stable imaging conditions necessary for atomic resolution.

Piezoelectric tuning forks have been employed as force sensors in the dynamic mode by many authors [4,5]. In this paper we report force distance studies on HOPG substrates at temperatures down to 1.7 K using a tuning fork sensor. Three methods for calibrating this sensor are compared in section 3. In section 4 we discuss simultaneous measurements of the force gradient and of the power dissipation vs. tip-sample separation.

## 2. Setup

### 2.1 The Microscope setup

We use a commercial cryo-SXM from Oxford Instruments which has been modified for working as a scanning force microscope utilizing a piezoelectric quartz tuning fork sensor for operation in the dynamic mode. The microscope is run in the He-gas atmosphere of the variable temperature insert of a cryostat at temperatures below 4.2 K and pressures of typically 5 millibar. The microscope is designed for the local investigation of semiconductor nanostructures in high magnetic fields, which requires a large scan range at low temperatures at the cost of lateral spatial resolution. Hence, atomic resolution cannot be achieved. Further details about the microscope can be found in Ref. 5.

### 2.2 Tuning fork and amplitude calibration

The tuning fork sensor is prepared by attaching a tungsten wire (diameter 13$\mu$m, lenght 300$\mu$m) at the end of one tuning fork prong and subsequent electrochemical etching of the wire to form a sharp tip with a tip radius of typically 10 nm as determined from transmission electron microscope images.

The tip is electrically isolated from the tuning fork contacts and separately connected to ground via a current preamplifier. On one hand this allows the measurement of the current, $I_T$, induced on the tip at constant tip-sample bias by the oscillating tip-sample capacitance. This capacitance oscillation is caused by the oscillation of the tip itself. On the other hand we can directly measure the tip-sample capacitance, $C(z)$, by applying an AC-voltage between the tip and the sample and ramping the tip-sample separation $z$.

The tuning fork is part of a phase-locked loop (PLL) which drives it always at its resonance frequency [5]. The PLL delivers the frequency shift of the tuning fork sensor and the dissipated power caused by tip-sample interactions.

In order to determine the mechanical and piezoelectric characteristics of the tuning fork sensor experimentally we have measured its mechanical vibration amplitude and its admittance as a function of frequency [6]. Applying a harmonic oscillator model we determined the effective mass of the oscillator $m = 0.33\text{mg}$, the mechanical stiffness $k = 15.4$ nN/pm and the piezoelectric coupling constant $\alpha = 4.2$ $\mu$C/m for room temperature operation [6]. At cryogenic temperatures these characteristic values will in general be different. However, we can argue that the effective mass of the oscillator should experience little change since the kinetic energy stored in the fork for a given tip speed at zero tip deflection will be barely changed. Using the above value for the effective mass we are then able to determine $k$ and $\alpha$ from the measurement of the admittance of the tuning fork at low temperatures alone, without the need of a simultaneous mechanical calibration of the amplitude. We will refer to this indirectly deduced low-temperature amplitude as the 'nominal amplitude' of the tip.

In our experiments below 4.5 K nominal tip oscillation amplitudes between 0.56 nm and 20 nm have been used. The quality factors of the oscillators ranged between 10000 and 120000 depending on temperature and pressure [6].

## 3. Force calibration

In order to translate measured frequency shifts $\Delta f$ into a quantity describing the tip-sample interaction we have calibrated our sensor. We use the force gradient approximation as the basic assumption, i.e.

$$\Delta f = \eta \cdot \frac{\partial F_z}{\partial z}, \quad (1)$$

where $(\partial F_z)/(\partial z)$ is the gradient of the tip-sample interaction force and $\eta$ is the proportionality constant which is to be determined by the calibration procedure. This approximation is valid for small tip-oscillation amplitudes. It is the limiting case of more general treatments discussed in the literature [7]. The calibration constant $\eta$ is independent of the details of the tip shape and depends only on the mechanical properties of the tuning fork. In the following we will compare three different methods of calibration and report the determined values of the calibration constant $\eta$.

*3.1 Harmonic oscillator model*

If we apply the harmonic oscillator model mentioned above for the resonance of our tuning fork, we can calibrate the sensor simply by determining the spring constant $k$ from the electrical resonance curve. We obtain the relation

$$\frac{\Delta f}{f} = \frac{(\partial F)/(\partial z)}{4k}$$

which differs by a factor of 1/2 from the result obtained for conventional cantilevers. This factor reflects the fact that only one prong of the fork senses the interaction but both prongs are oscillating. This method gives $\eta = f/(4k)$. For the sensor used for the studies described here we measured the resonance frequency $f = 32596.739$ Hz and the stiffness $k = 14.06$ nN/pm and therefore obtained a value of $\eta = 0.6$ Hz/(N/m). Since the resonance frequency of our sensors can be determined with an accurcacy better than 1 mHz the main source of error in this method may be found in the determination of $k$. It is calculated from $k = 4\pi^2 f^2 \cdot m$, thus any error in the assumption of an unaltered $m$ at low temperatures will cause an error in $\eta$.

*3.2 Calibration with the electrostatic force*

The two other calibration methods make use of the attractive electrostatic force acting between the tip and the sample if a constant tip-sample voltage $U$ is applied. It is given by the expression $F(z) = 1/2 \cdot (\partial C)/(\partial z) \cdot (U - \Delta(\mu_{ch}/e))^2$, where $C(z)$ is the tip-sample capacitance and $\Delta\mu_{ch}$ is the difference of the chemical potentials of the two materials which is usually taken to be equal to the work function difference between the two materials. Using eq. 1 we obtain for the frequency shift

$$\Delta f = \frac{\eta}{2} \cdot \frac{\partial^2 C}{\partial z^2} \cdot (U - \Delta(\mu_{ch}/e))^2 \qquad (2)$$

The measurement of $\Delta f(U)$ shown in Fig. 1 was performed at a large tip-sample separation and with a nominal tip-oscillation amplitude of 20 nm. For this tip-sample separation the electrostatic force is by far the dominant interaction. The data can be fitted extremely well by a parabola and therefore allows us to determine $\Delta\mu_{ch} = -64$ meV. From the measurement in Fig. 1 the calibration constant $\eta$ could be determined using eq. 2 if the second derivative of the capacitance were known at the same distance. For the determination of this quantity we have used two distinct techniques which will be described in the following.

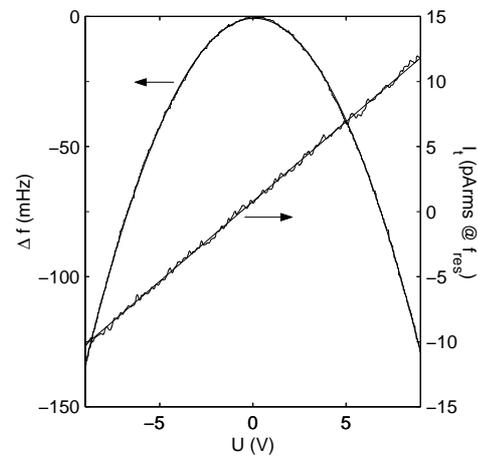

Figure 1: $\Delta f(U)$ measured at 4.5 K (left axis) and Kelvin current (right axis).

*3.2.1 Direct measurement of the capacitance*

We have measured the capacitance $C(z)$ depicted in Fig. 2 by applying an AC tip-sample voltage of 100mV(rms) at a frequency of 95 kHz. The measurement was fitted with a smooth analytical curve in order to allow a better determination of the second derivative. Inserting the result in eq. 2 we obtain $\eta \approx 1$ Hz/(N/m).

The accuracy of this method is mainly determined by the limited accuracy of the determination of the very small tip-sample capacitance which results in an uncertaintiy of the $\eta$-value of more than a factor of two for the measurement shown.

*3.2.2 Kelvin method*

The second approach for determining the second derivative of the capacitance employs the Kelvin method in which the gradient of the tip-sample capacitance is directly measured by oscillating the tip over the sample surface at a constant bias voltage and recording the current induced on the tip by the motion. This current is given by

$$I_T = 2\pi i f A \cdot \frac{\partial C}{\partial z} \cdot (U - (\Delta\mu_{ch})/e).$$

Fig. 1 shows an example of an $I(U)$ curve measured at the resonance frequency and a nominal tip-oscillation amplitude $A = 20$ nm for a large tip-sample distance. A linear fit to the curve allows the determination of $(\partial C)/(\partial z)$. The second derivative can be obtained by repeating this measurement as a function of $z$. Inserting the result into eq. 2 leads us to the calibration $\eta = 0.8$ Hz/(N/m). For this calibration method the tip-oscillation amplitude has to be known and any error in this number will lead to an error in $\eta$.

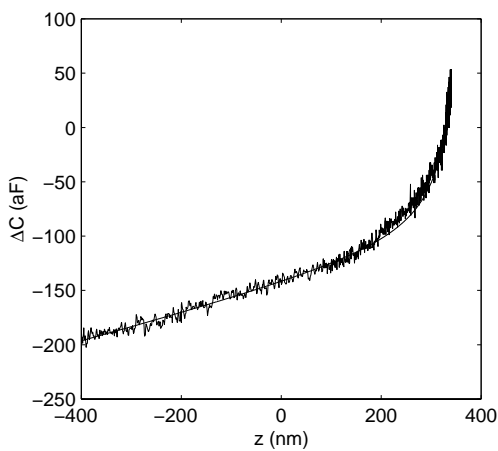

Figure 2: Tip-sample capacitance $C(z)$ as a function of the tip-sample separation $z$ measured at 1.7 K.

## 4. Force distance curves

Fig. 3a shows the frequency shift $\Delta f$ as a function of the tip-sample separation $z$ at zero bias voltage for different tip-oscillation amplitudes between 0.56 nm and 2.0 nm. All these curves have been measured for both sweep directions and no significant hysteresis was found. At large $z$, all the curves exhibit the well-known slow decrease of $\Delta f$ with decreasing $z$ until a minimum is reached. Beyond this minimum $\Delta f$ increases steeply with a rate of up to 10 Hz/Å at the smallest tip-oscillation amplitude. At a given $z$, larger amplitudes lead to smaller frequency shifts in accordance with the theoretical results in Ref. 7. The minimum in $\Delta f$ shifts to larger $z$ and becomes more shallow when the amplitude is increased. For the smallest amplitudes below 0.4 nm the curves are almost identical. We therefore conclude that for these amplitudes the force-gradient approximation is still valid even when the tip is very close to the sample.

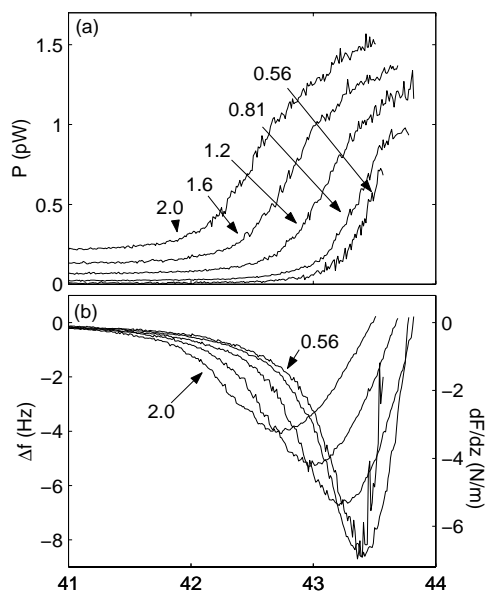

Figure 3: (a) Frequency shift vs. distance measured at 1.7 K for various tip-oscillation amplitudes. (b) Dissipated power for different amplitudes.

Fig. 3b shows the corresponding measurements of the dissipated power $P$ as a function of the tip-sample separation $z$. It is calculated from $P = U_{exc} \cdot I_{TF}$, where $U_{exc}$ is the excitation voltage applied to the tuning fork and $I_{TF}$ is the tuning fork current measured in phase with the excitation. We use the fact that the power dissipated in the system, in particular the power dissipated during tip-sample interaction, has to be supplied by the electrical excitation of the tuning fork. At large $z$ the dissipated power is constant. When the tip is approached to the surface the power starts to increase into the pW-range. For larger tip-oscillation amplitudes the onset of the dissipation is at larger tip-sample separations.

Even at the smallest tip-oscillation amplitudes the onset of the power dissipation occurs at a tip-sample separation at which the frequency shift is negative and the tip experiences an overall attractive force. We believe that at these tip-sample separations the foremost tip atoms start to come into intimate contact with the surface atoms and in addition to the conservative tip-sample interaction forces there occur dissipative phenomena. We speculate that phonon emission into the tip and the sample, or electrical losses due to the induced Kelvin current might be responsible for this effect. Further investigations are necessary to clarify the origin of the power dissipation in the system. However, it is interesting to notice that the power increase occurs at a position in the $\Delta f(z)$ curves where other authors have reported true atomic resolution [2,3]. At smaller tip-sample separations the tip must deform massively but the reproducibility of our $\Delta f(z)$ curves suggest that this deformation is elastic.

## 5. Conclusions

In this paper we have reported force-distance studies measured below 4.5 K using piezoelectric tuning fork sensors. Very high quality factors up to 120000 allow the detection of frequency shifts of less than 1 mHz. Small tip-oscillation amplitudes down to 0.5nm have been realized. Three methods for relating frequency shifts with force gradients led to similar calibrations of the force sensor. The setup performs well in magnetic fields up to 8 T and is therefore ideally suited for the future investigation of semiconductor nanostructures.

Measurements were performed in which the force gradient and the dissipated power were measured simultaneously as a function of the tip-sample separation on a HOPG surface. The data suggest that already at relatively large tip-sample separations when the tip experiences a net attractive force the foremost tip atoms touch the surface and power is dissipated. At smaller tip-sample separations the tip is significantly elastically deformed and at the same time the power dissipation increases strongly.

## Acknowledgements

We thank K. Karrai for very fruitful discussions on the use of tuning forks as force sensors and W. Allers for very stimulating discussions about force-distance studies. Financial support by ETH Zurich is gratefully acknowledged.